\documentstyle[12pt]{article}
\newcommand{\be}{\begin{equation}}   \newcommand{\ee}{\end{equation}}
\newcommand{\bd}{\begin{displaymath}} \newcommand{\ed}{\end{displaymath}}
\newcommand{\baa}{\begin{array}{lll}} \newcommand{\eaa}{\end{array}}
\newcommand{\ba}{\begin{eqnarray}}    \newcommand{\ea}{\end{eqnarray}}
\newcommand{\la}{\label}               
\newcommand{\Ds}{\displaystyle}  \newcommand{\q}{\bar q}

\begin{document}
{\flushright
     E2-94-482,
     JINR, Dubna}

 \vspace*{1cm}
\begin{center}
{\large \bf QCD SUM RULES FOR PION WAVE FUNCTION REVISITED}
\end{center}
\begin{center}
A.P. BAKULEV
\footnote{on leave of absence from VINITI RAN, Moscow, Russia;
E-mail bakulev@thsun1.jinr.dubna.su}
and
S.V. MIKHAILOV
\footnote{on leave of absence from Rostov State University, Rostov-Don,
Russia; E-mail mikhs@thsun1.jinr.dubna.su}\\
{\em Bogoliubov Laboratory of Theoretical Physics, JINR, Dubna, Russia}
\end{center}
\begin{abstract}
{\footnotesize
We analyze new QCD sum rules for the pion wave function (WF)
$\varphi_{\pi}(x)$, obtained recently in \cite{arnew}
in the non-local vacuum condensate method for non-diagonal correlators,
and suggest a new approach for extracting WF of the $\pi$-meson and the mass
and WF of the first resonance.
As a result, we obtain approximately the same form of the pion WF by two
different methods and for two different ansatzes of non-local quark
condensates. We predict the mass of the $\pi'$-resonance and obtain the form
of its WF.}
\end{abstract}

\section {Introduction}

An important problem in the theory of strong interactions is to
calculate, from the first principles of QCD, the distribution functions
$f_{p/H}(x)$~\cite{Feyn} and hadronic wave functions
 $\varphi_{\pi}(x) ,\ldots ~$ $ \varphi_N(x_1,x_2,x_3),...$~\cite{{CZ77}}
which  accumulate all the necessary information about non-perturbative
long-distance  dynamics of the theory.
These phenomenological functions appear naturally as a result of
applying ``factorization theorems" to hard inclusive and exclusive
processes \cite{CZ77}, \cite{ar77}, \cite{bl78}.

   Another kind of such phenomenological numbers are the quark and gluon
condensates
$\langle:\q(0) q(0):\rangle, ~~\langle:G(0)G(0):\rangle \ldots$,
the basic parameters of the QCD SR approach \cite{svz}
reflecting the non-perturbative nature of the QCD vacuum.
A usual practice is to calculate the hadronic functions
$f(x),~~\varphi(x)$ by using the condensates as input parameters  \cite{cz82}.

  The situation with hadronic WF's is more complicated than with
distribution functions: the firsts appear only in an integrated form
like some convolutions. It seems now that only the QCD Sum Rule (SR) approach
and lattice calculation \cite{lat91} can provide an information about the form
of hadronic WF.
The most popular set of hadronic WF, due to ~V.~L.~Chernyak,
A.~R.~Zhitnitsky and I.~R.~Zhitnitsky (CZ) \cite{cz82}, was produced
with the help of QCD SR for the first moments of WF's. These SR were
based on the diagonal correlator of appropriate axial currents and
the condensates of the lowest dimensions; the pion WF thus obtained
has the well-known ``two-hump'' form. But now it is known that the
hadronic functions are rather sensitive to the structure of
non-perturbative vacuum \cite{nlc}. Therefore one should use
a non-local condensate like $\langle:\q(0) E(0,z) q(z):\rangle$
which can reflect the complicated structure of the QCD vacuum.
(Here $E(0,z)=P\exp(i \int_0^z dt_{\mu} A^a_{\mu}(t)\tau_a)$ is
the Schwinger phase factor required for gauge invariance.)

Earlier, one of the authors (S.~M. together with A.~Radyushkin)
constructed a modified SR with non-local condensates
and demonstrated that the introduction of the correlation length $\Lambda$
for condensate distributions produces much smaller values for the
first moments of pion WF than the CZ values \cite{nlwf89},~\cite{nlwf92}.
This leads to the form of the pion WF strongly different from the CZ
form and close to the asymptotic form  $\varphi_{\pi}^{as}(x) = 6x(1-x)$.
Our goal in the present paper is to obtain {\bf directly the form} of
the pion WF and the first resonance, using the available ansatz for
non-local condensates (section 2). This program has been suggested
recently and realized in \cite{arnew}, we develop alternative methods of
extracting wave functions from this sum rule for a non-diagonal correlator.
We suggest two different ways: the first uses integrated (in a sense)
properties of ansatzes (see section \ref{Sect3})
and may be considered as an alternative to traditional Borel SR;
while the second uses their local properties (see section \ref{Diff})
and follows to the Borel SR idea. The results obtained by these
independent methods agree with each other.

\section {Non-diagonal correlators and pion WF}

\subsection{The models of non-local quark condensates}

 It is appropriate to recall some general features
of derivation of the QCD sum rules with non-local condensates
{}~\cite{nlc},~\cite{nlwf89},~\cite{nlwf92},~\cite{arhq}.
The quark and gluon fields are taken in the Fock--Schwinger gauge
$A_{\mu}(z)z^{\mu} = 0$, therefore path-ordered
exponentials $E(0,z)=P\exp(i \int_0^z dt_{\mu} A^a_{\mu}(t)\tau_a)$
are equal to $1$.
It is convenient to parametrize the  $z^2$-dependence of the simplest
bilocal  quark condensate
$\langle:\bar q(0)q(z):\rangle \equiv \langle:\bar q(0)q(0):\rangle Q(z^2)$
in analogy with the $\alpha$--representation of a propagator
\footnote{ In deriving these sum rules we can always make a Wick
rotation, i.e., we assume that all coordinates are Euclidean, $z^2 <0$.}
\be
Q(z^2) =
\int_{0}^{\infty} e^{s z^2/4}\, f(s)\, ds .
      \label{eq:qq}
\ee
The  correlation  function $ f(s)$ may be interpreted as
``the distribution function of quarks in the vacuum'' ~\cite{nlwf89}
since its  $n$th moment is proportional to the
matrix  element of the local operator with  $D^2$ to $n$th power:
\be
\int_{0}^{\infty} s^n f(s)\,  ds = \frac{1}{\Gamma(n+2)}
\frac{\langle:\bar q (D^2)^n q :\rangle}
{\langle:\bar q  q:\rangle}.
\ee
For the lowest two moments one obtains
\ba \la{eq:norm}
\int_{0}^{\infty}  f(s)\,  ds =1;~~~
\int_{0}^{\infty} s f(s)\,  ds = \frac1{2} \frac{\langle:\bar q D^2 q:\rangle}
{\langle:\bar q  q :\rangle} \equiv \frac{\lambda_q^2}{2},
\ea
with $\lambda_q^2$  meaning the average virtuality of vacuum quarks.

  In a similar way, parametrizing the quark-gluon non-local condensate
$$\langle: \bar q(0) ig(\sigma G(0)) q(z): \rangle \equiv
\langle: \bar q(0) q(0): \rangle Q_1(z^2),$$
one can introduce the quark-gluon distribution function $f_1(s)$.
(Note the approximate character of this definition: its l.h.s. appears
as a three-point correlator and only after reducing to two-point geometry
can be parametrized in terms of two-quark non-local
condensates~\cite{nlwf89}.)

  To  construct  models of  non-local condensates, one should satisfy
some constraints. For instance, if we assume vacuum matrix elements
$\langle:\bar q (D^2)^{k} q :\rangle$ to exist, then the function $f(s)$
should decay faster than $1/s^{k+1}$ as $s \to \infty$.
If all such matrix elements exist (for all $k$), a possible choice could be
a function $ f(s) \sim e^{-s^2/\sigma^2}$,
or $ f(s) \sim e^{-s/\sigma}$, $etc.$ at large $s$.
The opposite, small-$s$ limit of $f(s)$  is determined by the
large-$|z|$ properties  of the function $Q(z^2)$.
This behavior has been analyzed in detail in ~\cite{arhq} in the framework of
QCD SR for heavy quark effective theory (HQET): it has been demonstrated
that for a large Euclidean $z$, non-local quark condensate
$Q(z^2) \sim  e^{- |z| \Lambda}$ with $\Lambda = (M_Q-m_Q)|_{m_Q\to \infty}$
being the lowest energy level of the mesons in HQET
(numerically, $\Lambda$ is around 0.45 $GeV$). This means that
$f(s) \sim e^{-\Lambda^2/s}$  in the small-$s$ region.
Another hint to prove this behavior of $f(s)$ can be obtained from the
results of the lattice calculation \cite{lat84}
where the exponential decay behavior for the
correlator $\langle: G(0)\tilde E(0,x)G(x):\rangle$ of the gluon vacuum
strengths has been found. For this reason one should expect in a theory
with confinement
\ba \la{ans1}
\langle:\q(0)q(z):\rangle \mathop{\sim}_{z^2 \to -\infty} \exp(-|z|\Lambda)
\ea
where $\Lambda$ is the correlation length. This law generates the
above-mentioned behavior for $f(s)$ at small $s$.
So, we arrive at the set of ansatzes AI:
\ba
\mbox{A1}:\,\,\,\,f(s) &=& N_1
\cdot\exp\left(-\Lambda^2/s-s^2 \sigma_1^2\right)
\quad\!\,\,\,\,\, \mbox{--- Gaussian decay}
\nonumber \\
\mbox{A2}:\,\,\,\,\,f(s) &=& N_2
\cdot\exp\left(-\Lambda^2/s - s \sigma_2\right)
\quad\,\,\,\,\,\,\, \mbox{--- exponential decay}
\\
\mbox{A3}_k:~f_k(s) &=&  \frac{\exp\left(-\Lambda_k^2/s\right)}{\Gamma(k-1)}
\frac{1}{s} \left(\frac{\Lambda_k^2}{s}\right)^{(k-1)}
\quad  \mbox{--- inverse power decay} \nonumber
\ea
where $ \Lambda^2 = 0.2 \, GeV^2$ and $ \Lambda_k^2 =
 \Lambda^2 (k-2)$,
the normalization constants $N_i$  and the
$\sigma_i$-parameters are fixed by  eq.(\ref{eq:norm}), where for  the average
virtuality of vacuum quarks we take the
usual QCD sum rule value \cite{belioffe} $\lambda_q^2 \simeq 0.4 \,
GeV^2$.

   Ansatz A1 has been used in \cite{arnew}, it is effectively close to
ansatz A2. The second ansatz leads to the same  behavior for the
$\langle :\q(0) q(z):\rangle$ as the massive causal scalar propagator
with a shifted argument does:
\ba \la{A2}
 \langle:\q(0)q(z):\rangle \mathop{\sim}_{z^2 \to - \infty} N_2\cdot
{\sqrt{\pi \Lambda}\over\left(\sigma_2+|z^2|/4\right)^{3/4}}
\exp\left(-2\Lambda\sqrt{\sigma_2+|z^2|/4}\right) \\
\Ds N_2 = {1\over 2\frac{\Lambda}{\sqrt{ \sigma_2}}
K_1\left(2\Lambda \sqrt{\sigma_2} \right)},
{}~~K_1(z) \mbox{ is the modified Bessel function}. \nonumber
\ea

    The ansatz A3 has analytic properties in $s$ different from those of A2,
is very convenient for calculations and imitates the
$\delta(s-\Lambda)$ when the parameter $k \gg 1$. But it doesn't look
sufficiently ``physical'' due to the absence of higher local
condensates mentioned above. Nevertheless, we shall test it also at
the values $k=3,~4,~5$. This ansatz leads to the asymptotic behavior
\ba\la{A3}
 \langle:\q(0)q(z):\rangle \mathop{\sim}_{z^2 \to -\infty}
\frac{\sqrt{\pi}}{\Gamma(k-1)}
  \left(\frac{z\Lambda}{2}\right)^{\frac{2k-3}{2}}
\exp\left(-\Lambda |z|\right).
\ea

\subsection{Sum rule}
\label{SR}
   Let us write down the model sum rule for the ``axial'' wave
functions $\varphi_{\pi^{\ldots}}(x)$ of the pseudoscalar mesons
 suggested in ~\cite{arnew}:
\ba \la{a17}
\varphi_{\pi}(x) &+& \varphi_{\pi'}(x) e^{-m_{\pi'}^2/M^2}
+\varphi_{\pi''}(x) e^{-m_{\pi''}^2/M^2} + \ldots
\equiv  \Phi(\frac{1}{M^2},x)  \\
&=& \frac{M^2}{2}\left (1-x+\frac{\lambda_q^2}{2M^2} \right) f(xM^2)
+ (x \to 1-x) \nonumber
\ea
with the function $f(s)$ specified in the preceding section.
It should be noted that this sum rule results from the approximations
(for a detailed discussion, see ~\cite{arnew}):
\begin{enumerate}
\item reduction of three-point correlators to two-point ones;
\item $f_1(s)=2\lambda_q^2\cdot f(s)$;
\item contributions of higher non-local condensates
$\langle \bar q G G q \rangle, \langle \bar q G G G q \rangle, \ldots$
could be neglected.
\end{enumerate}

   As has been shown in~\cite{arnew} the function $\Phi(\frac{1}{M^2},x)$,
{\em i.e.}, the weighted sum of all WF's, is given by two humps
(one centered at $x_A=s_A/M^2$, where $s_A$ is the point of maximum for
the ansatz distribution function $f(s)$, and the other -- at $x_A=1-s_A/M^2$)
moving as $M^2$ changes. When $M^2$ increases, the humps become
narrower, higher and more close to boundary points $x=0$ or $x=1$.
For $M^2=1\,~GeV^2$, {\em e.g.},  the function  $\Phi(\frac{1}{M^2},x)$
looks very much like the CZ WF (see Fig.\ref{fig-AR}).
However, $\Phi(\frac{1}{M^2},x)$ is not just the pion WF:
the larger $M^2$, the larger contamination from higher states.
Moreover, the (``axial'') WF of the higher pseudoscalar mesons
$\pi'$,  $\pi'' , \ldots$ produce zero total integrals
(whereas the pion WF is normalized to unity, see also ~\cite{cz93})
and therefore they should oscillate (see, e.g., Fig.\ref{fig-A2-res}.)

  At low $M^2$, the pion WF dominates in the total sum
$\Phi(\frac{1}{M^2},x)$ (however, one cannot take too low $M^2$ because
the operator product expansion fails for $M^2 < \lambda_q^2$).
When $M^2 = 0.4 \,~GeV^2$, it was observed that $\Phi(\frac1{M^2},x)$
is very close to the asymptotic wave function of the pion
(see Fig.\ref{fig-AR}).

  In our opinion, this very nice picture of successive switching on
resonances has an essential drawback:
there is no strict criterion for selecting the value of $M^2$ for
determining WF of the ground state (pion). The author of~\cite{arnew}
writes: "$\ldots$ one should be more accurate here, since even the modest
increases of $M^2$ to $0.5 \,~GeV^2$ or $0.6 \,~GeV^2$ induce humps
in $\Phi(x,M^2)$" (see Fig.\ref{fig-AR}).

\section {Method of Integral Projector}
\la{Sect3}

     As we have seen before, all the resonances contribute
to $\Phi(\frac1{M^2}, x)$ and the problem is to distinguish between the
ground state ($\varphi(x)$) and resonances $(\varphi_i(x))$ and
between resonances. We should know the whole spectrum $\mu_i$
to extract the ground state.
We represent here a method to obtain $\varphi_i(x)$ and $\mu_i$
{\bf step by step}.
It is based on an integral transformation of equation (\ref{a17}).

As a result, we shall obtain approximately the same form
for the ground state and the first resonance by using essentially
different ansatzes A2 and A3.

\subsection {General formulas}

  In what follows it's convenient to use a new variable
$\tau \equiv 1/M^2$ instead of the Borel parameter $M^2$.
The r.h.s. of SR (\ref{a17}), i.e. $\Phi(\tau,x)$ is defined for
$\tau \in \big[0, \frac{1}{\lambda_q^2}\big]$. Let us consider this function
in the whole complex plane of $\tau$ and define the projector
$P(N, \tau_0, \omega)$:
\ba   \la{2}
P(N, \tau_0, \omega)\big(\cdots\big)  =  \frac{N!}{2\pi i \omega^N}
\int_{C} \frac{\exp(\omega (\tau-\tau_0))}{(\tau-\tau_0)^{N+1}}
\big(\cdots\big) d\tau~,
\ea
where the contour $C$ is a vertical line $C=(c-i\infty,~c+i\infty)$ with
$c > 1/\lambda_q^2$, the $c$ lyes righter than any pole of the integrand,
{}~~$\omega > 0$ and $\tau_0\in(0,\frac{1}{\lambda_q^2}\big]$.
To obtain the result of this projector action on the initial
SR-representation
\be \la{3}
 \varphi(x)+ \sum_{i=1} \varphi_i(x) \exp(-\tau \mu_i)  = \Phi(\tau,x)  ,
\ee
consider its action on a simple exponential. Evidently, by
closing the contour $C$ to the left we get (due to the well-known
residues theorem):
\be   \la{4}
P(N, \tau_0, \omega) \exp(-\mu \tau) = \theta(\omega-\mu) \cdot
\exp(-\mu\tau_0) \left(1-\frac{\mu}{\omega}\right)^N.
\ee
Then, for $\omega > 0$, we have:
\ba \la{5}
\varphi(x)+ \sum_{i\ge 1}\theta\left(\omega-\mu_i\right)\varphi_i(x)
\left(1-\frac{\mu_i}{\omega}\right)^N\exp\left(-\tau_0\mu_i\right)
 = P(N, \tau_0, \omega) \Phi(\tau, x) \equiv \bar \Phi_N(\omega,x).
\ea
To understand this result, it is instructive to consider the case
$N=0, \ \tau_0=0 $:
\be \la{6}
\varphi(x)+ \sum_{i\ge 1}^{\mu_i < \omega} \varphi_i(x) =
P(0, 0, \omega) \Phi(\tau, x)
\ee
We see that by varying $\omega$ one can switch on (switch off) more and more
resonances in the l.h.s. of SR.
For $N>0$ one can successively determine the positions (masses $\mu_i$)
of resonances. Formula (\ref{5} ) is the main result of this subsection.

  Consider now the meaning of the introduced parameters of the projector
$P(N,\tau_0,\omega)$:
\begin{enumerate}
\item The parameter $\omega$ has a clear physical meaning --- it produces a
{\bf real division} of the set of resonances ( with $\mu_i > \omega$).
Contrary to the role of the ``unphysical'' Borel parameter $M^2$
that supplies an exponential cut of higher resonances, it gives a
(smoothed by powers $\Ds\left(1-\frac{\mu}{\omega}\right)^N$) real step cut.
\item The real life is more complicated than the simple model (\ref{6}):
to suppress the non-desirable asymptotic behavior of the function
$\Phi(\tau,x)$ for $\tau \rightarrow \infty$ (in order to neglect the
integral over a semicircle closing the initial contour $C$), one should use
the parameter $N > 0$. Besides, the projected SR with different $N$
can be used for determining the subsequent resonances.
\item At last, $\tau_0$ defines the significant region of variable $\tau$
which mainly contributes to the integral.
\end{enumerate}

Let us consider the third feature in detail.
Note that the physical meaning has only large-$|z|$ asymptotics of
the non-local quark condensate $\langle:\q(0)q(z):\rangle$.
But no information about this behavior can be extracted from the
known values of the first few moments. Contrary to that,
the behavior of the condensate at small $|z|$ isn't fixed at all,
and different kinds of the behavior of the correlation function $f(s)$
are allowed at a large $s$ (or a small $\tau$).
The reason is the dominance of perturbative contribution in this region.
Different ansatzes A1, A2, A3 have been suggested earlier in this way.
It is easy to show that (\ref{ans1}) generates the exponential behavior
of the correlation function $f(s)$ at a small $s$,
$f(s) \sim \exp(-\Lambda^2/s)$ that corresponds to the law
$$ \Phi(\tau,x) \sim \exp(- \tau \Lambda^2)$$
at a large $\tau$ for the r.h.s. of SR.

  As the behavior of $\Phi(\tau)$ at small $\tau$ isn't fixed
we always choose the value of $\tau_0$ in some region near point
$1/\lambda_q^2$. Since the final result, in any way, should not depend on
the behavior of $\Phi(\tau)$ at a small $\tau \ll 1/\lambda_q^2$,
{\bf this is a constraint on the available correlation function $f(1/\tau)$
and the value of $\tau_0$}. The practical recipe is that one takes
the parameter $\tau_0$ so that the contribution of the region near point
$\tau=0$ to the integral (\ref{2}) must be small
(see the integration with ansatz A2 in the next subsection).

 Consider now as an illustration, the result of the action of
$P(N,\tau_0,w)$ on the $n$-th term of the operator product expansion
(OPE) $\langle O_n\rangle \cdot \tau^n$ in the r.h.s. of the traditional SR,
proportional to $\tau^n$.
After explicit integration we obtain for $N > n$:
\ba \la{8}
  P(N,\tau_0,\omega) \tau^n = \tau_0^n \cdot
  \left( \sum_{k=0}^{N} C^k_N \frac{\Gamma(n+1)}{\Gamma(n+1-k)}
  \frac1{(\omega\tau_0)^k}\right) =
  \tau_0^n \cdot\left(1+\frac{N n}{\omega \tau_0}+\ldots \right)
\ea
One can see the re-formation of the usual set of power corrections:

 In the limiting case  $\omega \tau_0 \to \infty$  (at a fixed $\tau_0$),
when all the resonances are included into the l.h.s.,
the projector $P(N,\tau_0,\omega)$  is similar to the projector on
a single value of $\tau$, i.e. its kernel in the integral representation
is close to $\delta(\tau - \tau_0)$:
  $$P(N,\tau_0,\omega) \tau^n \approx \tau_0^n ,$$
and the SR is restored again to its initial form (with a new argument
$\tau \to \tau_0$).

For intermediate values of $\omega$, the cut of the tail of resonances
in the l.h.s. (i. e. those which satisfy $\omega < \mu_i$) leads to the
relative growth of theoretical contributions in the r.h.s. of
the SR (see (\ref{8}))
with respect to the remaining resonance contributions.
In other words, it can be interpreted as an effective increase in
the scale of the standard vacuum expectation values $\langle O_n\rangle$.
This example demonstrates, in particular, the strong dependence
of SR treatment on the adopted model for the phenomenological part of SR.

\subsection{How to extract a ground state WF}

   Let us assume that the experimental mass values are known
$\mu_0 =0,~ \mu_1=1.7,~ ... $  and they correspond to ansatzes A2 or A3.
Then, we can obtain an expression for $\varphi(x)$ by employing
equation (\ref{5}) with the parameter $\omega=\omega_1 \leq \mu_1$
that corresponds to the saturation of the ground state $\varphi(x)$
\be \la{9}
\varphi(x)=P(N,\tau_0,\omega_1)\Phi(\tau,x)=\bar\Phi_N(\omega_1,x).
\ee
Then, the final result should not depend on the parameter $N$ strongly.
However, in the real life we don't know the position $\mu_1$ exactly,
therefore we may get a contamination from the next state.
This contamination becomes smaller when $N$ increases, due to the power
suppression near threshold $\mu_i$. So, the following criterion
of saturation of the ground state can be suggested:
the form of the curve $\varphi(x)=\bar\Phi_N(\omega_1,x)$  doesn't
depend on $N$ at some $\omega_1$.
\noindent This is a very simple but no so effective criterion for the
following reason: the region of comparatively small $\omega$
(i. e. $\omega \le 1~ GeV^2$) seems to be transient,
here the corresponding WF is not well normalized (its normalization is of
the order of $0.5$ as compared to needed $1$).
The normalization of WF restores at $\omega \ge 1.5~ GeV^2$,
but the correction for the first resonance is too small to be extracted
with the help of this criterion.

  To integrate the r.h.s. of SR (\ref{9}), we take residues:
there are two residues in the case of ansatz A2 at the points $\tau=\tau_0$
and $\tau=0$ (essential singularity). At the end we obtain that
the saturation realizes near $\omega =1.8 \div 2.0 ~ GeV^2 $ and
the function $\bar \Phi_N(\omega,x)$ doesn't change noticeably for
$N=0,~ 1,~2,~3$. The parameter $\tau_0$ is chosen in accord with
the suggested criterion of smallness of the contribution from the residue
at $\tau=0$ and appears to be $\tau_0 \sim 2.5 ~GeV^{-2}$.
The curve is in form rather close to the asymptotical WF, and
the second moment of WF $\langle \xi^2\rangle=0.19$
(the corresponding curves are shown in Fig.\ref{fig-A2-123}).

   We have completely different behavior of the r.h.s. of SR near
$\tau \sim 0$ for ansatz $A3_k$, $\Phi(\tau,x) \sim \tau^k$.
For this reason, we are should use the projector with the parameter $N=k+2$
for this ansatz. We proceed here with ansatzes with $k = 3, 4, 5$.
The integral in the r.h.s., $\bar\Phi_N(\omega_0,x)$, is determined by
the one residue at the point $\tau=\tau_0$. Performing the procedure
like for ansatz A2 with the same value of $\tau_0$, we obtain
slightly thinner forms of $\varphi(x)$ for all these ansatzes.
As it will be clear later, the most adequate to pion physics is ansatze
$A3_k$ with $3\le k\le 4$. For this case we observe the saturation of
the r.h.s. of SR at $\omega =\omega_0 \approx  2~ GeV^2$ with
$\langle\xi^2\rangle \approx 0.16$.
(We could conclude about this saturation by the smallness of
variations of $\bar\Phi_N(\omega_0,x)$ with $N$,
of course the WF normalization now reaches unity.)
The results are also close to the asymptotic WF, see Fig.\ref{fig-A3}.

\subsection{How to extract resonance WF's}

Suppose we know the values of $\mu_1,~\mu_2,~\ldots$, then we can extract
the first resonance by inverting equation (\ref{5}), e.g., at $N=1$:

\ba \la{10}
\Ds \varphi_1(x) = \frac{\left(\bar\Phi_1(\omega,x)-\bar\Phi_1(\mu_1,x)\right)}
{1-\frac{\mu_1}{\omega}}\exp\left(\tau_0\mu_1\right)
\ea
where $\mu_1 < \omega \leq \mu_2$ corresponds to saturation
of the first resonance.

The later equation may be improved if one takes into account a more
detailed spectrum model. Let us suggest that the resonance contribution
to the spectral density has a finite width $2\Delta$, e.g.:
$$ \delta(s-\mu)\cdot\varphi_1(x) \to \rho(s-\mu, \Delta)\cdot\varphi_1(x),~
  \mbox{ where } \rho(s-\mu, \Delta)\mid_{\Delta \to 0}~=\delta(s-\mu)$$
This leads to a more complicated expression in the l.h.s. of SR (\ref{a17})
than simple exponentials. Applying the projector $P(N, \tau_0, \omega)$
to the l.h.s. of SR with this new spectral density $\rho$
one can easily obtain
\be   \la{4a}
P(N, \tau_0, \omega) \int_0^{\infty} e^{-s \tau} \rho(s-\mu, \Delta) ds=
 \left(1 + \frac{d}{\omega d \tau_0}\right)^N
 \int_0^{\infty} \theta(\omega-s) e^{-s \tau_0} \rho(s-\mu, \Delta) ds
\ee
instead of expression (\ref{4}). To estimate the main effect of the
finite width $\Delta$, we apply the simplest (and rather rough)
``step-model'' for  $\rho$
\be \la{spmod}
 \rho(s-\mu, \Delta) = \theta(\mu+\Delta > s) \theta(s > \mu-\Delta)
\cdot \frac{1}{2 \Delta},
 \ee
at natural constraint $\Delta \ll \mu$. Substituting (\ref{spmod}) into
(\ref{4a}) and integrating we obtain a new expression for the l.h.s. of SR.
The latter leads to modified equation for $\varphi_{1}(x)$;
at $N=1$ and $\omega > \mu + \Delta$ we arrive at the expression
\ba \la{10mod}
\Ds \varphi_1(x) = \frac{\left(\bar\Phi_1(\omega,x)-
\bar\Phi_1(\mu_1,x)\right)}
{1-\frac{\mu_1-\Delta \mu_1}{\omega}}\exp\left(\tau_0\mu_1\right)
\cdot \frac{\Delta\tau_0}{\sinh(\Delta\tau_0)},~~
\Delta\mu_1=\Delta\cdot\left(\coth(\Delta\tau_0)-\frac{1}{\Delta\tau_0}\right)
\ea
The final numerical result due to finite width in (\ref{10mod}) is not
large for parameters of the first resonance ~\cite{data}
$\Delta \approx 0.18~GeV^2,~\mu_1 \approx 1.7 GeV^2$
but may be important for the next resonance.

  Let us assume now that we don't know the spectrum $\{\mu_i\}$.
How can we find a resonance for the particular ansatz?
Evidently, equation (\ref{5}) has the best sensitivity to it at
the lowest $N$, because the threshold effects are suppressed for large $N$.
However SR with $N=0$ is not also good:
there is a too rough simulation of the continuous (over $\omega$) r.h.s.
by the set of step-like contributions.

  The case of SR with $N=1$ is more attractive:
here steps are smoothed by switching-functions
$\left(1-\frac{\mu_i}{\omega}\right)$ and the discontinuity is present only
in derivatives of the l.h.s. of SR. As it has been mentioned in
the discussion of the criterion for selecting resonances in the former
subsection, the magnitude of this discontinuity is too small to be trigger
resonances. If we examine the difference
$\bar\Phi_N(\omega,x) - \bar\Phi_{N+1}(\omega,x)$, then we obtain the
very convenient criterion for resonances:
\be \la{11}
\theta\left(\omega-\mu_1\right)\varphi_1(x)\cdot\exp\left(-\mu_1\tau_0\right)
\frac{\mu_1}{\omega}\left(1-\frac{\mu_1}{\omega}\right)^{N}=
\bar\Phi_N(\omega,x) - \bar\Phi_{N+1}(\omega,x).
\ee
As one can see, at $\omega = \omega_{*} = \mu_1$, this difference is equal
to zero and for $\omega > \mu_1$ becomes negative, if $\varphi_1(x)<0$,
otherwise, positive, with $1/\omega$-decreasing at large $\omega$.
For ansatz A2 we see in Fig.\ref{fig-A2-dif} that the zero of this
difference at $N=1, ~x=0.5$ is reached for $\omega_{*} = 1.8~ GeV^2$
that is quite reasonable for the pion case
(the experimental value is $\mu_{1 exp} \approx 1.7\pm 0.25~ GeV^2$).
It should be emphasized that $\bar\Phi_1(\omega,x)$ imitates the
l.h.s. of  SR  as a function of $\omega$ rather well
despite of their different origin. Note here that the position of
the root $\omega_{*}$ depends also on the value of $x$:
in the region $0.35\le x\le 0.65$ this dependence is rather weak.

    So, we have found the position of the first resonance $\mu_1$
that agrees with $\mu_{1 exp}$, based on ansatz A2.
This is an unexpectedly good result for our rather crude model of
non-local condensates (see subsection \ref{SR}).
The saturation by the first resonance is reached near
$\omega \approx 2.5 \div 3 ~GeV^2$. The curves corresponding to
the resonance $\varphi_1(x)$ (for different values of $\omega$ near
$\omega=3~GeV^2$) are shown on Fig.\ref{fig-A2-res}.

   The same program for ansatz A3 with $k = 3, 4, 5$ produces the roots
$\omega_{*}(k) = 1.3,$ \, $3.0,$ \, $5.5 ~GeV^2,$ respectively.
For this reason the choice $3 \le k \le 4$ is more adequate to
the pion case. Here we also observe the saturation of normalizations and
forms of the ground state WF before switching on the first resonance
(i.e. at $\omega < \omega_{*}(k)$). The curves of the ground state WF
corresponding different $k = 3, 4, 5$ are very close.

\section {Method of Differential Projectors}
\la{Diff}
\vspace*{-2mm}
\subsection{General formulae}
We shall construct here a trick to extract the ground state and resonances,
based on the criterion traditional for QCD SR approach:
a physical quantity, $\varphi(x)$, should not depend on the Borel parameter
$M^2$ on the stability plateau.
However, as we have seen in section 2, the weighted sum of all WF
in the initial representation (\ref{a17}) has a strong dependence
on $M^2$ even for $M^2 \leq 1~GeV^2$ due to contamination by the resonances,
and there is no stability of the form of a curve at all.

  Let us suggest that the positions of the first $n$ resonances are
known exactly. Then, one may hope to obtain the ground state $\varphi(x)$
that doesn't depend on $M^2$ in the region [$\lambda_q^2, M^2_n$],
by cancelling the contributions of these resonances.
The contamination in this case will be defined by the contributions of
the $(n+1)$-th resonance and subsequent resonances and will be suppressed
by the exponential factor $\sim \exp( -\mu^2_{n+1}/M^2)$.
\footnote{ It is similar to the formulation of the ``inverse problem'':
to obtain the $\varphi(x)$ one uses the known spectrum $\{\mu_i\}$. }
If the positions of some resonances are not well known,
one can vary the parameters $\mu_i$ to obtain the best fit of stability.
Of course, this proposal will ``work'' if the resonances
are well distinguished and the corresponding pre-exponential factor
$\varphi_{n+1}(x)$  is not too large (for  any $x$).
Note at least that the renorm-group evolution of the condensates
with respect to $M^2$ should be taken into account to complete this
analysis in a separate publication.

To cancel the contribution of a set of resonances, we shall use a
differential operator. Apparently, one can obtain for one resonance
$\varphi_i$ ( here $D \equiv \partial_{\tau}$ ):
$$\left(1+{D\over \mu_i}\right) e^{-\tau\mu_i} \varphi_i(x)=0;~~
\left(1+{D\over\mu_i}\right)\varphi(x)=\varphi(x).$$
 From this result it is easy to find the action of the projector $Q_0(n)$
\footnote{ We are grateful to A.~V.~Radyushkin who suggested
us to use just the same form of the projector}
\ba\la{12}
Q_0(n) \equiv \prod^n_{i=1}\left(1+{D\over\mu_i}\right),
\ea
on a finite set of the resonances $L_{n} \equiv
 \sum^n_{i=1}\exp(-\mu_i\tau)\varphi_i(x)$:
\ba \la{13}
Q_0(n)\left(\sum^n_{i=1} e^{-\mu_i\tau}\varphi_i(x)\right)=0;~~
Q_0(n)\left(\varphi(x) + \sum^n_{i=1}
e^{-\mu_i\tau}\varphi_i(x)\right)
=\varphi(x).
\ea
So, applying the projector $Q_0(n)$ to eq. (\ref{3}), one obtains
the expression for the ground state $\varphi(x)$
\ba\la{14}
\varphi(x)=Q_0(n)\Phi(\tau,x) + O(e^{-\mu_{n+1}\tau}).
\ea
We can extend the region of stability for $\varphi(x)$ with respect to $\tau$
(or $M^2$) by adding subsequent resonances to projector $Q_0(n)$
(and making the last term in eq.(\ref{14}) still smaller).

    By a similar procedure, the projector for extracting the $k$-resonance
>from the set $L_{n}$ could be obtained:
\ba \la{15}
Q_k(n) =  e^{\mu_k\tau} \cdot\prod^n_{i\ne k}
\left(1+{\mu_k+D\over\mu_i-\mu_k}\right).
\ea
The contamination of the $\varphi_k(x)$ resonance could be more strong
than for $\varphi(x)$ due to the factor $\exp(\mu_k\tau)$.

    It would be worthwhile to rewrite expression (\ref{14}) as a form of
Laplace representation:
\ba\la{16}
\varphi(x)=\int^{\infty}_{0}e^{-p\tau}\prod^N_{i=1}
\left(1-\frac{p}{\mu_i}\right) \tilde\Phi(p,x) dp;~\mbox{ where }~
\Phi(\tau,x) = \int^{\infty}_{0} e^{-p\tau}\tilde\Phi(p,x) dp
\ea
and then to employ the identity
\ba\la{17}
\prod^n_{i=1}\left(1-\frac{p}{\mu_i}\right)\equiv
\exp\left(-ps_1-p^2\frac{s_2}{2}-\ldots-p^n\frac{s_n}{n}-\ldots\right),~
\mbox{ where } s_k\equiv\sum^n_{i=1}\left(\frac1{\mu_i}\right)^k .
\ea
Now, all the information about the spectrum {\bf is accumulated in
$s_i$-parameters} in the final formula:
\ba\la{18}
\varphi(x)=\int^{\infty}_0 \exp
\left(-p\tau -ps_1 -\sum^{\infty}_{k=2}p^k\frac{s_k}{k}\right)
\tilde\Phi(p,x)~dp + O(e^{-\mu_{n+1}\tau}).
\ea
Expression (\ref{18}) is the main result of this subsection:
it is very convenient for applications of any models of
the spectrum $\{\mu_i\}$, and allows one to use different approximations.

\subsection{The spectrum model and the ground state WF}

We suggest here the following spectrum model:\\
{\em The positions of the first two resonances are $\mu_1 = 1.7,~
\mu_2 \approx 3 ~(GeV^2)$ in accord with experimental data  and
estimations of subsection 3.3.
For subsequent $\mu_i$ we adopt the ``physically appealing'' oscillator's
model with a finite spectrum:
 $\mu_3=\mu,~ \mu_4=\mu+1, \ldots~ \mu_n= \mu+n-3$.}

  Now we rewrite expression (\ref{18}) in a form without the first two
resonances, and all the model suggestions concentrate in the intergrand
exponential:
\ba\la{19}
\varphi(x)=\left(1+\frac{D}{\mu_1}\right) \left(1+\frac{D}{\mu_2}\right)
\int^{\infty}_0
\exp\left(-p\tau-ps_1-\sum^{\infty}_{k=2}p^k\frac{s_k}{k}\right)
\tilde\Phi(p,\tau) dp + O(e^{-\mu_{n+1}\tau}),
\ea
where
$$s_k=\sum^n_{i=3}\left(\frac1{\mu_i}\right)^k. $$
Our model allows one to apply any approximation to the integral in the r.h.s..
We shall not here describe the details (see Appendix A). It results in a
``three-resonance formula'' for expression (\ref{19}) with a shifted
argument $\tau$
\ba\la{20}
\varphi(x) \approx \Phi_{3R}(x,\tau) \equiv
\left(1+ \frac{D}{\mu_1} \right ) \left(1+ \frac{D}{\mu_2} \right )
\left(1+ \frac{D}{\mu} \right ) \Phi(\tau,x) \mid_{\tau \to \tau + s_1},
{}~~ s_1= \sum_{i=3}^{n} \frac{1}{\mu_i}.
\ea
Note here that the shift $\tau \to \tau+s_1$ effectively reduces
the available region of the parameter $M^2$:
$$ M^2 \to \frac{M^2}{1 + s_1M^2}. $$
Evidently, this ``scaling'' effect improves the stability of $\varphi(x)$
with respect to $M^2$.

\begin{figure}[p]
  \caption{A2-ansatz: $\Phi_{3R}(x,\frac{1}{M^2})$ with
  $M^2 = 0.5 ~GeV^2$(line), $M^2 = 1.0 ~GeV^2$(small-dashed),
  $M^2 = 1.5 ~GeV^2$(big-dashed)  vs. asymptotic pion wave function (dotted)}
     \label{fig-A2-DfM}
       \end{figure}
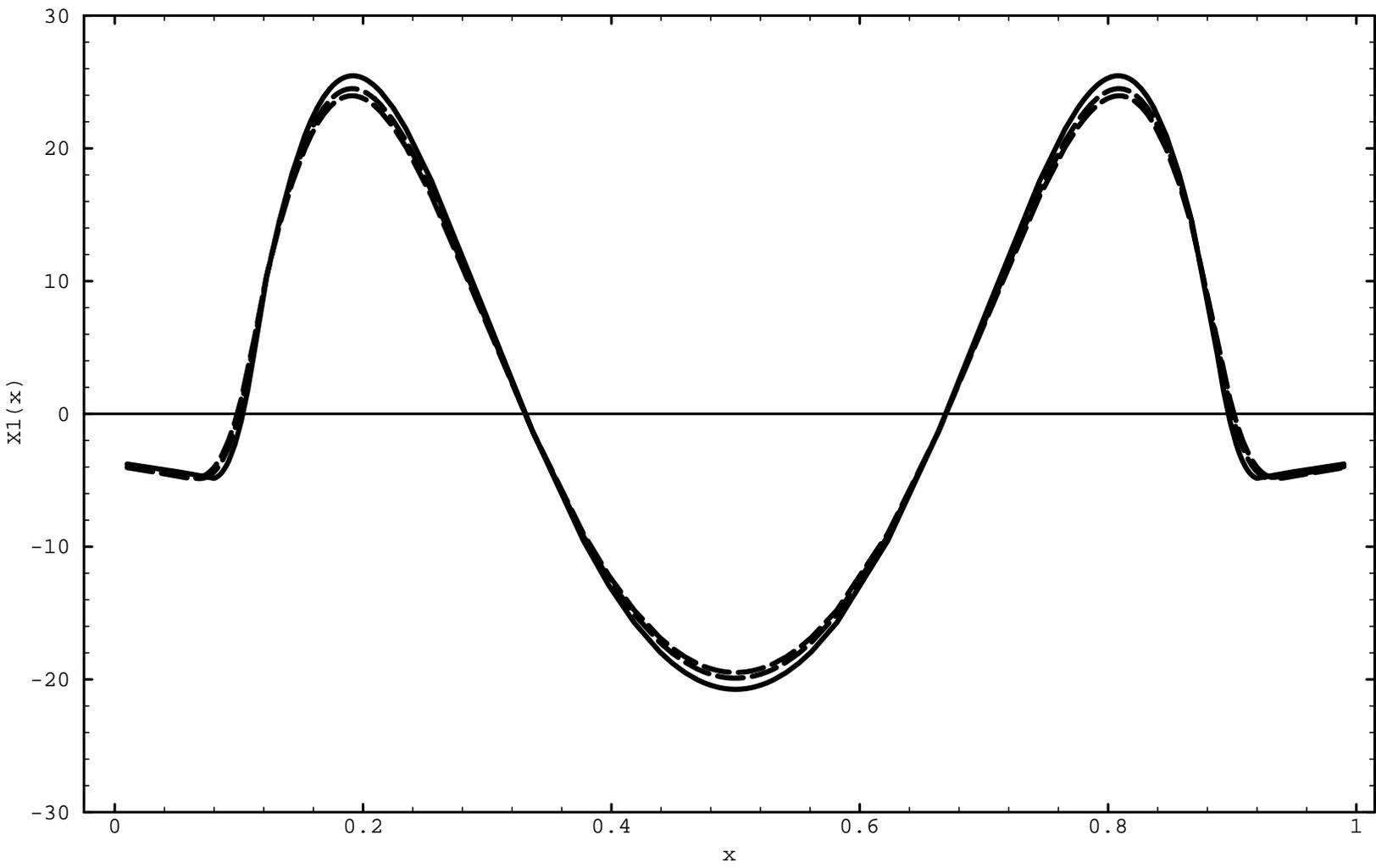
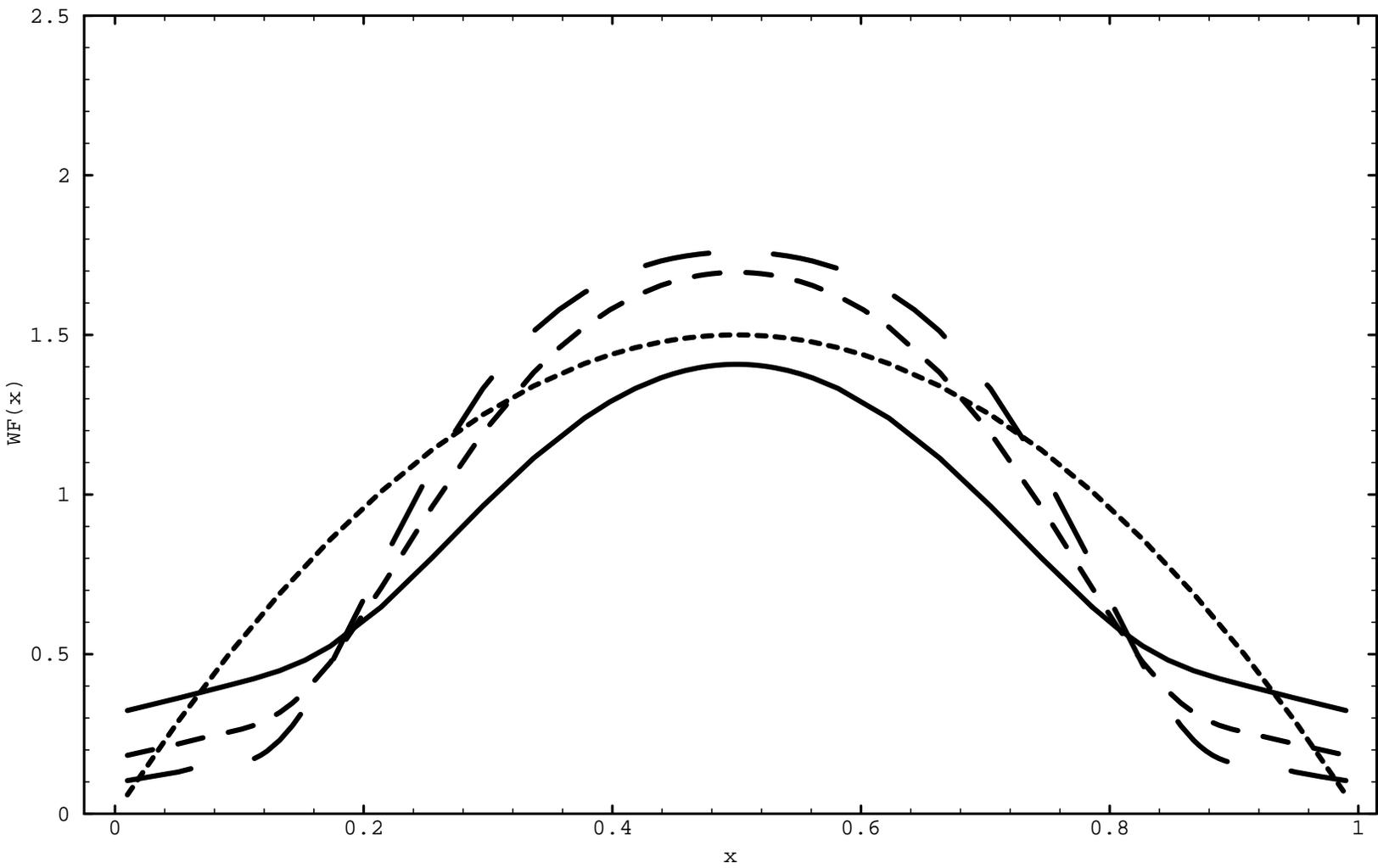

The ``reduced'' expression (\ref{20}) depends on two parameters,
$\mu$ and $n$. We can vary them to obtain the best region of stability
for $\varphi(x)$ with respect to $M^2$. The best fit leads to the values:
$\mu \approx 4~GeV^2,~~ n=11$ for  ansatz A2. For these values
of parameters the form of the curve $\varphi(x)$ doesn't depend on
$M^2$ visibly in the region    $M^2 \in (0.5 - 2)~ GeV^2$,
and has a rather slow evolution when $M^2$ increases.
The form of curve (in the stability region) is similar to the asymptotic
WF again (see Fig.\ref{fig-A2-DfM}),
and the second moment of WF $\langle\xi^2\rangle=0.16$ for $M^2=1~ GeV^2$.
So, we may conclude that the results of the "differential method"
confirm the results for pion WF obtained by the "integration method"
in Sec. \ref{Sect3}.

\section{Conclusion}
In this paper, we have considered a model sum rule for the pion wave
function and pseudoscalar resonances based on the non-diagonal
correlator \cite{arnew}. The theoretical side (r.h.s.) of this sum rule
depends only on the non-local condensate. We suggest two different ways
to extract the wave functions of the ground and resonance states of
the meson from the sum rule.
Two kinds of ansatzes with different analytic properties for the
condensate distribution have been used for final numerical estimates.
We have obtained qualitatively similar results for the pion
wave function by using both these methods and different condensate ansatzes:
their forms are close to the asymptotic wave function. Moreover,
we obtained the shape of the wave function and the position of
the first resonance near
$m_1^2 \approx 1.8 ~~GeV^2, (m_{1 exp}^2 = 1.7~~GeV^2)$,
based on the first kind ( A2 ) of the ansatz. Our results confirm
the qualitative conclusions about shapes of the wave function of pion and
the resonance obtained in ~\cite{arnew}.

\vspace*{5mm}

{\large \bf Acknowledgments}\\ \vspace{2mm}

We are grateful to ~A.~V.~Radyushkin  for clarifying criticism.
His paper ~\cite{arnew} inspired the present investigation.
We also thank R.~Ruskov for fruitful discussions of the main results.
We are indebted to the International Science Foundation
(grant RFE-000) and to the Russian Foundation for Fundamental Research
(contract 93-02-3811) for financial support. One of us (M.S.)
is grateful to Prof. A. Di Giacomo for fruitful discussions
and  warm hospitality at the Physics Department of Pisa University.

\newpage
\appendix
\vspace{3cm}
{\Large\bf Appendix}
\section{Three-resonance formula}
\setcounter{equation}{0}
\renewcommand\theequation{\thesection{.\arabic{equation}}}

The  oscillator spectrum model leads to the evident expression for
coefficients $s_k$ in (\ref{17}):
\be\la{a1}
 s_k = \sum_{i=0}^{n-\mu} \frac{1}{(\mu + i)^{k}} = \zeta(k; \mu) -
  \zeta(k; n+1).
\ee
To  derive the three-resonance formula (\ref{20}) we shall use
a set of successive approximations (\ref{a2} - \ref{a5}).
Substituting (\ref{a1}) into the identity (\ref{17}) we obtain:
\ba\la{a2}
\sum^{\infty}_{k=2}p^k\frac{s_k}{k} \equiv
 \sum^{\infty}_{k=2}p^k\frac{\left(\zeta(k; \mu) - \zeta(k; n+1)\right)}{k}
{}~~ \mathop{\approx}^{n \gg 1} \nonumber \\
\approx \left[\mu\cdot (1-z_1)\ln(1-z_1)-(n+1)\cdot (1-z_2)\ln(1-z_2)\right]
 \mid_{z_1 \le 1},
\ea
where $z_1=p/\mu$, $z_2=p/(n+1)$.
\ba\la{a3}
\Ds \exp
\left[-\left( \mu\cdot(1-z_1)\ln(1-z_1)-n\cdot(1-z_2)\ln(1-z_2)\right)\right]
{}~ \mathop{\approx}^{n \gg 1}~
  {e^{-p}\over \left(1-\frac{p}{\mu}\right)^{(\mu-p)}}.
\ea
Simple numerical analysis demonstrates the validity of the approximate
equation for the r.h.s. of (\ref{a3}):
\ba\la{a4}
\Ds \theta(\mu-p) {e^{-p}\over \left(1-\frac{p}{\mu}\right)^{(\mu-p)}}
 \approx \theta(\mu-p)\left(1-\frac{p}{\mu}\right).
\ea
The main condition for the next approximation is that the Laplace
representation $\tilde\Phi(p,\tau) $ for A2 is concentrated on
a bounded support of about few $GeV^2$, and the integral in (\ref{18})
over $p$ is saturated in the interval $(0, \mu)$. So, we may hope that
\ba\la{a5}
\int_{0}^{\infty} dp \left[ (1-\frac{p}{\mu}) \theta(\mu-p) \right]
\tilde\Phi(p,x) \exp\left(-p\tau -ps_1\right)
\approx \left(1+\frac{D}{\mu}\right) \Phi(\tau, x) \mid_{\tau \to \tau+s_1}.
\ea
Substituting the set of approximations (\ref{a2} - \ref{a5}) into
expression (\ref{19}), we arrive at three-resonance formula for
$\varphi(x)_{\pi}$
\ba\la{a6}
\varphi(x)_{\pi} \approx
\left(1+ \frac{D}{\mu_1} \right ) \left(1+ \frac{D}{\mu_2} \right )
\left(1+ \frac{D}{\mu} \right ) \Phi(\tau,x) \mid_{\tau \to \tau + s_1}.
\ea

\newpage
\begin{figure}[p]
  \caption{$\Phi(x,M^2)$ with
  $M^2 = 0.4 ~GeV^2$(line), $M^2 = 0.6 ~GeV^2$(small-dashed),
  $M^2 = 1.0 ~GeV^2$(big-dashed)  vs. asymptotic pion wave function (dotted)}
     \label{fig-AR}
       \end{figure}
\begin{figure}[p]
  \caption{A2-ansatz: $\bar\Phi_N(\omega,x)$ with $N = 1$(line),
  $N = 2$(small-dashed), $N = 3$(big-dashed)  vs. asymptotic pion
  wave function (dotted)}
     \label{fig-A2-123}
       \end{figure}
\begin{figure}[p]
  \caption{A3$_4$-ansatz: $\bar\Phi_N(\omega,x)$ with  $N = 6$(line),
  $N = 7$(small-dashed) vs. asymptotic pion wave function (dotted)}
     \label{fig-A3}
       \end{figure}
\begin{figure}[p]
  \caption{A2-ansatz: $\bar\Phi_1(\omega,0.5)-\bar\Phi_2(\omega,0.5)$}
     \label{fig-A2-dif}
       \end{figure}
\begin{figure}[p]
  \caption{A2-ansatz: $\varphi_1(x,\omega)$ with
  $\omega = 2.5 ~GeV^2$(line), $\omega = 3.0 ~GeV^2$(dashed) and
   $\omega = 3.5 ~GeV^2$(dotted)}
     \label{fig-A2-res}
       \end{figure}
\end{document}